\documentclass{elsart}
\usepackage{epsfig}
\usepackage{rotate}
\usepackage{amssymb}
\usepackage{amsmath}

\newcommand\taumu{\tau \to \mu \nu \nu}
\newcommand\mutau{\mu \to \tau}

\begin{document}

\begin{flushright}
{\large
INR-RAS/2001-024}
\end{flushright}

\vspace{1.5cm}

\begin{frontmatter}
\begin{center}
{\large \bf Probing lepton flavour violation in\\
$\nu_{\mu} + N \to \tau + ...$ scattering and $\mutau$ conversion\\
 on nucleons.} 
\end{center}
\vspace{.5cm}

\begin{center}
S.N.~Gninenko\footnote{E-mail address:
 Sergei.Gninenko\char 64 cern.ch}, M.M.~Kirsanov,
N.V.~Krasnikov and V.A. Matveev\\
\vspace{.5cm}
{\it Institute for Nuclear Research of the Russian Academy of Sciences,\\
Moscow 117312}
\end{center}


\begin{abstract}
We study lepton flavour-violating  interactions which could result in the 
 $\tau$-lepton production in the $\nu_{\mu}N$ scattering or
in $\mu \to \tau$ conversion on nucleons at high energies.\ 
 Phenomenological bounds on the strength of  
$\bar{\tau}\nu_{\mu}\bar{q}q^{'}$ interactions are extracted  
from the combined result of the NOMAD and CHORUS experiments on searching for
 $\nu_{\mu} \to \nu_{\tau}$ oscillations. 
Some of these bounds supersede limits from 
rare decays.\\ 
We also propose  a ``missing energy'' type  experiment searching for 
 $\mu \to \tau$ conversion  on nucleons. The experiment
 can be performed at a present accelerator or at a future neutrino factory.
\end{abstract}
\end{frontmatter}

\section{Introduction}

In the standard model of strong and electroweak interactions (SM) 
with massless neutrinos lepton flavour is conserved. However many extensions 
of the SM including the simplest extension with massive neutrinos predict 
flavour-violating interactions. For instance leptoquarks which occur 
naturally in all unified models such as $SU(5)$ \cite{1} or 
Pati-Salam $SU(4)$ \cite{pati} , in models with quark-lepton substructure 
\cite{buch1} and in supersymmetric models with R-parity violation 
\cite{barger}. 
In general leptoquarks have baryon or lepton number violating couplings 
and they must be very heavy in order to avoid rapid proton decay or large 
Majorana neutrino masses. However, in theories with conserved baryon and 
lepton number leptoquark masses and couplings satisfy much weaker bounds 
which come mainly from the analysis of flavour-violating processes. 
Many searches for specific reactions, namely $\mu - e$ conversion on nuclei
 \cite{wintz},  $\mu \rightarrow e \gamma$ 
\cite{mega} and $\mu \rightarrow 3e$ \cite{bell} decays, 
rare $K$ \cite{kaon}, $D$- $B$- \cite{pdg} and 
$\tau-$ decays \cite{cleo} and also search for lepton 
flavour violation in $e-p$ collisions \cite{zeus} and neutrino interactions
\cite{nutev}
have been performed resulting in stringent limits on the coupling strength 
of lepton flavour violating four-fermion $\bar{l}l^{'}\bar{q}q^{'}$ 
interactions.\ 

There are also projects to improve some of these bounds 
significantly, see e.g. \cite{psi,meco} including experiments at a
neutrino factory.\ The neutrino factory is a
future powerful tool for investigation of the nature of neutrino,
in particular neutrino osccilations.\ It could also 
be useful for investigation of other physics beyond the standard model, 
see e.g. \cite{nufact} and references therein.\ This facility would be 
an intense source of muons and might enable to improve the current upper 
limits on flavour-violating transitions, involving in particular 
$\mu$'s or $\nu_{\mu}$'s, by several orders of magnitude. 

In this note we have  suggested to probe lepton flavour violation 
through  $\tau$-lepton production in  $\nu_{\mu}N$ scattering or 
$\mu + N \to \tau + N$ conversion on nucleons at high energies.\ It 
should be stressed that experiments on  search for $\mu$-$\tau$ conversion 
will probe flavour-violating interactions, in particular like 
$ L = \bar{\mu}\tau[G_1\bar{u}c + G_2\bar{d}b + G_3\bar{s}b]$. 
Due to trivial kinematical constraints 
(e.g. $\tau \to \mu + D$ decay cannot occur, 
because of $m_{\tau} < m_{\mu} + m_D$), there are no stringent bounds 
on such interaction from the existing
 experimental bounds on rare 
$\tau$-lepton, such as $\tau \to \mu + \gamma$ \cite{cleo}, 
and charm/beauty meson 
decay rates \cite{pdg}.\ For investigation of such  
flavour-violating interactions to perform  
experiments on searching for $\mu$-$\tau$ conversion is important.\       

This paper is organised as follows.\ In Sec.2 we extract 
 phenomenological bounds on the 
strength of  $\bar{\tau}\nu_{\mu}\bar{q}q^{'}$ interactions   
from the current experimental bounds on search for $\nu_{\mu} - 
\nu_{\tau}$ oscillations from the  NOMAD and CHORUS experiments.\  
These bounds can compete from  bounds derived 
from $\tau$-decays.\ In Sec.3 we consider a "missing energy" type experiment
to search for lepton flavour violation in $\mu - \tau$ conversion
on nucleons at high energies.\ In Sec. 4 we present our conclusions.

\section{Phenomenological bounds}

The latest combined limit of the CHORUS and NOMAD experiments on the 
probability of $\nu_{\mu} \rightarrow \nu_{\tau}$ oscillations is \cite{petti}:  
 
\begin{equation}
P(\nu_{\mu} \rightarrow \nu_{\tau}) < 0.6\cdot 10^{-4}
\end{equation}

Using the bound of Eq.(1) one  can find that 
NOMAD data lead to the bound 
\begin{equation}
P(\nu_{\mu} \rightarrow \tau) = 
\frac{\sigma(\nu_{\mu} N \rightarrow \tau +...)}
{\sigma(\nu_{\mu} N \rightarrow  \mu + ...)} < 0.6 \cdot 10^{-4}
\end{equation}
on flavour-violating cross section of $\nu_{\mu}$ neutrino with isoscalar 
nucleon target. 

Bound of Eq.(2) allows to obtain restriction on flavour-violating interactions
$\bar{\tau}\nu_{\mu}\bar{q}q^{'}$ which lead to nonzero cross section for 
the flavour changing reaction $\nu_{\mu} N \rightarrow \tau + ...$. 
Let us consider  the case of scalar leptoquarks. $SU(3) \otimes 
SU(2) \otimes U(1)$ invariant Lagrangian of scalar and vector leptoquarks 
satisfying baryon and lepton number conservation has been written down in 
ref. \cite{okun}. For us the following interactions leading to 
flavour-violating 
reaction $\nu_{\mu} N \rightarrow \tau + ...$ are interesting:
\begin{eqnarray}
&&L^{'} = (g_{1L}\bar{q}^c_Li\tau_{2}l_L + g_{1R}\bar{u}^c_Re_R)S_1 +
(h_{2L}\bar{u}_Rl_{L} +h_{2R}\bar{q}_{L}i\tau_2e_R)R_2 + h.c. \\
 \nonumber
&&-m^2_1 S^{+}_1S_1 -m^2_2R^{+}_2R_2
\end{eqnarray}
Here $q_L$, $l_L$ are the left-handed quark and lepton doublets, and $e_R$, 
$d_R$, $u_R$ are the right-handed charged leptons, down- and up-quarks, 
 respectively; $\psi^c = C\bar{\psi}^T$ is a charge-conjugated fermion 
field. The subscripts L, R of the coupling constants denote the lepton 
chirality. The indices of the LQ's give the dimension of their $SU(2)$ 
representation. Colour, weak isospin and generation(flavour) indices have 
been suppressed. After integration over heavy leptoquark fields one can 
obtain effective flavour-violating four-fermion interactions
\begin{eqnarray}
&&L_{4\Delta F} = \frac{g_{1L}^2}{m^2_1}(\bar{q}^c_Li\tau_2l_L)
(\bar{q}^c_Li\tau_2l_L)^{+} + 
\frac{g_{1L}g_{1R}}{m^2_1}(\bar{q}^c_Li\tau_2l_L 
\bar{e_R}u^c_R) \\ \nonumber 
&&+ ... + \frac{h_{2L}h_{2R}}{m^2_2}(\bar{u}_Rl_L\bar{e}_R
i\tau_2q_L) + ...
\end{eqnarray}
Especially interesting are couplings with the first and second quark 
generations:
\begin{eqnarray}
&&L^{'}_{4\Delta F} = \frac{g_{1L,1}g_{1L,2}}{m^2_1}(\bar{s}^c_L\nu_{{\mu}L} - 
\bar{c}^c_L\mu_L)(\bar{\nu_{\tau}}d^c_L - \\ \nonumber
&&\bar{\tau}_{L}u^c_L) + 
\frac{g_{1L,2}g_{1R,2}}{m^2_1}(\bar{s}_L^c\nu_{\mu L} \bar{\tau}_Rc^c_R - 
\bar{c}_L^c \mu_L \bar{\tau}_Rc^c_R)  + \\ \nonumber
&&\frac{g_{1L,1}g_{1R,2}}{m^2_1}(\bar{d}_L^c\nu_{\mu L}\bar{\tau_R}c^c_R - 
\bar{u}_L^c\mu_L\bar{\tau}_Rc_R^c) +
\frac{h_{2L,2}h_{2R,1}}
{m^2_2}(\bar{\tau}_Rd_L\bar{c}_R\nu_{\mu L} - 
\bar{c}_R\mu_{L}\bar{\tau}_Ru_L) \\  \nonumber 
&&+ \frac{h_{2L,1}h_{2R,2}}{m^2_2}(\bar{u}_R\nu_{\mu,L}\bar{\tau}_Rs_L - 
\bar{u}_R\mu_L\bar{\tau}_Rc_L) + h.c. 
\end{eqnarray}

For the couplings of Eq.(5) bounds resulting from flavour-violating 
$\tau$-lepton 
decays bounds are absent.\ So, the CHORUS and NOMAD data are in fact 
the single ones that
allow to extract the most stringent bounds on couplings of 
four-fermion interactions of Eq.(5). In quark-parton 
model  the cross section of the inclusive $\nu_{\mu} N \rightarrow 
\mu +...$ scattering on the 
isoscalar nucleon is determined by the formula 
\footnote{Here we use notations of ref. \cite{buch2}} 
\begin{eqnarray}
&&\sigma(\nu_{\mu}N \rightarrow \mu+..) \equiv \frac{1}{2}(\sigma(\nu_{\mu}p 
\rightarrow \mu +..) + \sigma(\nu_{\mu}n 
\rightarrow \mu + ...)) =  \\ \nonumber 
&&= \frac{G^2s}{2\pi}((Q + S) +\frac{1}{3}
(\tilde{Q} - \tilde{S})) \approx 
\frac{G^2s}{\pi}\cdot 0.22
\end{eqnarray}
where \cite{buch2}
\begin{equation}
Q = U +D + S \equiv \int^1_0 [u(x) +d(x) +s(x)]dx \equiv \int^1_0 q(x)dx 
\approx 0.44
\end{equation}
and $G$ is the Fermi constant.

\begin{eqnarray}
&&\tilde{Q} = \tilde{U} + \tilde{D} + \tilde{S} \equiv  
\int^1_0[\tilde{u}(x) 
+ \tilde{d}(x) + \tilde{s}(x)]dx =  \\ \nonumber
&& \int^1_0\tilde{q}(x)dx \approx 0.07
\end{eqnarray}
\begin{equation}
\tilde{U} \approx \tilde{D} \approx 0.03
\end{equation}
\begin{equation}
S \approx \tilde{S} \approx 0.01
\end{equation}
\begin{equation}
U \approx 0.28
\end{equation}
\begin{equation}
D \approx 0.15
\end{equation}
Using the four-fermion flavour-violating interaction of Eq.(5) and 
quark-parton model one can find that 
\begin{eqnarray}
&&\sigma(\nu_{\mu}N \rightarrow \tau + ...) = 
\frac{s}{96\pi}[\frac{g_{1L,1}g_{1L,2}}{m^2_1}
(3S + \frac{1}{2}(\tilde{D} + \tilde{U})) \\ \nonumber
&& +\frac{g_{1L,2}g_{1R,2}}{m^2_1}(3S)
+\frac{g_{1L,1}g_{1R,2}}{m^2_1}(\frac{U + D}{2}) + \\ \nonumber
&& \frac{h_{2L,2}h_{2R,1}}{m^2_2}(\frac{3}{2}(U + D)) +  
\frac{h_{2L,1}h_{2R,2}}{m^2_2}(\tilde{S} + 
\frac{3}{2}(\tilde{U} + \tilde{D}))]
\end{eqnarray}

From formulae of Eqs.(6-13) we find bounds

\begin{equation}
\frac{g_{1L,2}g_{1L,2}}{m^2_1} < 0.4 \cdot G
\end{equation}
\begin{equation}
\frac{g_{1L,2}g_{1R,2}}{m^2_1} < 0. 16 \cdot G
\end{equation}
\begin{equation}
\frac{g_{1L,2}g_{1R,1}}{m^2_1} < 0.08 \cdot G
\end{equation}
\begin{equation}
\frac{h_{2L,2}h_{2R,1}}{m^2_2} <  0.03 \cdot G
\end{equation}
\begin{equation}
\frac{h_{2L,1}h_{2R,2}}{m^2_2} < 0.16 \cdot G
\end{equation}

Note that in supersymmetric models with R-parity violation 
\cite{buch2} one of the 
possible R-violating terms in the superpotential has the form
\begin{equation}
W^{'} = g_{1L}l_{L}i\tau_{2}q_Ld^c_R
\end{equation}
After integration over right-handed down squarks we find in 
particular the effective 
interaction 
\begin{equation}
L_{\Delta F}^{'} = \frac{g_{1L,12}g_{1L,23}}{m^2_{sq}}
(-\bar{u}^c_L\mu_{L} +\bar{d}^c_L\nu_{\mu,L})\cdot 
(-\bar{c}^c_L\tau_{L} + \bar{s}_L^c\nu_{\tau})^{+} + ...
\end{equation}
which in fact coincides with the first term of the interaction (5). 
Here $m_{sq}$ is the mass of right-handed down-squark. So using the 
previous results we find that 
\begin{equation}
\frac{g_{1L,23}g_{1L,12}}{m^2_1} < 0.08\cdot  G
\end{equation} 

For   $g_{1L,23}g_{1L,12} =1$ we find that $m_{sq} \gtrsim 1000~GeV$.

One can extract bounds on effective coupling constants of the 
general four-fermion interaction describing the transition 
$\nu_{\mu}N \rightarrow \tau + ...$ \footnote{Here we consider only 
interactions of the first and second quark generations with 
$\nu_{\mu}$  and $\tau$}

\begin{eqnarray}
&&L_4 = \frac{4}{2^{1/2}}[G_{1,11}\bar{u}_R\nu_{\mu}\bar{\tau}_{R}d_L 
+  G_{1,12}\bar{u}_R\nu_{\mu}\bar{\tau}_{R}s_L +    
G_{1,21}\bar{c}_R\nu_{\mu}\bar{\tau}_{R}d_L+
G_{1,22}\bar{c}_R\nu_{\mu}\bar{\tau}_{R}s_L+ \\ \nonumber
&&G_{2,22}\bar{s}_L^c\nu_{\mu}\bar{\tau}_{L}c_L^c+ +
G_{2,21}\bar{s}_L^c\nu_{\mu}\bar{\tau}_{L}^cu_L^c +
G_{2,12}\bar{d}_L^c\nu_{\mu}\bar{\tau}_{L}c_L^c +
G_{2,11}\bar{d}_L^c\nu_{\mu}\bar{\tau}_{L}u_L^c + \\   \nonumber
&&G_{3,11}\bar{d}_L^c\nu_{\mu}\bar{\tau}_{R}u_R^c +
G_{3,12}\bar{d}_L^c\nu_{\mu}\bar{\tau}_{R}c_R^c +
G_{3,21}\bar{s}_L^c\nu_{\mu}\bar{\tau}_{R}u_R^c +
G_{3,22}\bar{s}_L^c\nu_{\mu}\bar{\tau}_{R}c_R^c + h.c.]
\end{eqnarray}

In quark-parton model for the interaction of Eq.(22) the 
$\nu_{\mu} \rightarrow \tau$ inclusive cross secrion is 
determined by the formula
\begin{eqnarray}
&& \sigma(\nu_{\mu}N \rightarrow \tau + ...) = \frac{s}{12\pi}[
G^2_{1,11}(\frac{3}{2}(\tilde{U} + \tilde{D}) + \frac{1}{2}(U+D)) +  
G^2_{1,21}( \frac{1}{2}(U+D)) + \\ \nonumber  
&&G^2_{1,12}(\frac{3}{2}(\tilde{U} + \tilde{D}) + S) +
G^2_{1,22}(S) + 
(G^2_{2,11}+G^2_{3,11})
(\frac{3}{2}(\tilde{U} + \tilde{D}) + \frac{1}{2}(U + D)) + \\ \nonumber  
&&(G^2_{2,21}+G^2_{3,21})
(3S + \frac{1}{2}(U+D)) +  
(G^2_{2,12}+ G^2_{3,12})( \frac{3}{2}(\tilde{U}+\tilde{D})) + 
(G^2_{2,22}+G^2_{3,22})(3S) ]
\end{eqnarray}  
Using NOMAD bound of Eq.(3) and  Eq.(6) for inclusive $\nu_{\mu} \rightarrow
\mu$ cross section we find bounds on parameters of flavour-violating 
four-fermion interaction of Eq.(24)
\begin{equation}
(G_{1,11}, G_{2,11}, G_{3,11}) < 3.3 \cdot 0.8\cdot 10^{-2}G
\end{equation}
\begin{equation}
G_{1,21} < 2.8 \cdot 10^{-2}   G
\end{equation}
\begin{equation}
G_{1,12} < 5.5 \cdot 10^{-2}  G
\end{equation}
\begin{equation}
G_{1,22} < 13.0 \cdot 10^{-2}  G
\end{equation}
\begin{equation}
(G_{2,12}, G_{3,12}) < 2.6 \cdot 10^{-2}G
\end{equation}
\begin{equation}
(G_{2,12}, G_{3,12}) < 7.5 \cdot 10^{-2} G
\end{equation}
\begin{equation}
(G_{2,22}, G_{3,22}) < 7.5  \cdot 10^{-2}  G
\end{equation}

It should be noted that in the search for $\mu$-$\tau$ conversion we probe in 
particular four-fermion flavour changing interactions. As it has been 
mentioned in the introduction bounds on the flavour changing interactions 
\begin{equation}
\Delta L = \bar{\mu}\tau[G_1\bar{u}c + G_2\bar{b}d + G_3 \bar{s}b]
\end{equation}
are not very stringent due to trivial kinematical reasons. For, instance 
$\tau$ lepton cannot decay to D-meson and muon, since 
$m_{\tau} < m_{\mu} + m_D$). So, bounds from $\tau$ lepton 
decays do not lead to bounds on $G_i$. Due to the same reason 
bounds from rare D- and B-mesons are also not very stringent. The 
study of $\mu$-$\tau$ conversion will allow to obtain stringent bounds on 
coupling constants $G_i$.

\section{Experimental search for the  $\mu \rightarrow \tau$ conversion.}

Here we describe the search for flavour-violating interactions of 
the $\mu\bar{\tau}\bar{q}q^{'}$ type by using the muon to tau-lepton  
conversion on nucleons at high energies.
 Note that the four-fermion interaction 
$\mu\bar{\tau}\bar{c}u$ does not contribute to $\tau$-decay, since 
$D$-meson mass is bigger than $\tau$-lepton mass, so for such interaction 
the flavour-violating bounds resulting from $\tau$-decay modes are not 
applicable, hence the search for $\mu - \tau$ conversion is in fact the 
single way to probe such interactions.

\begin{figure}
\centering
{\hspace{-.5cm}\epsfig{file=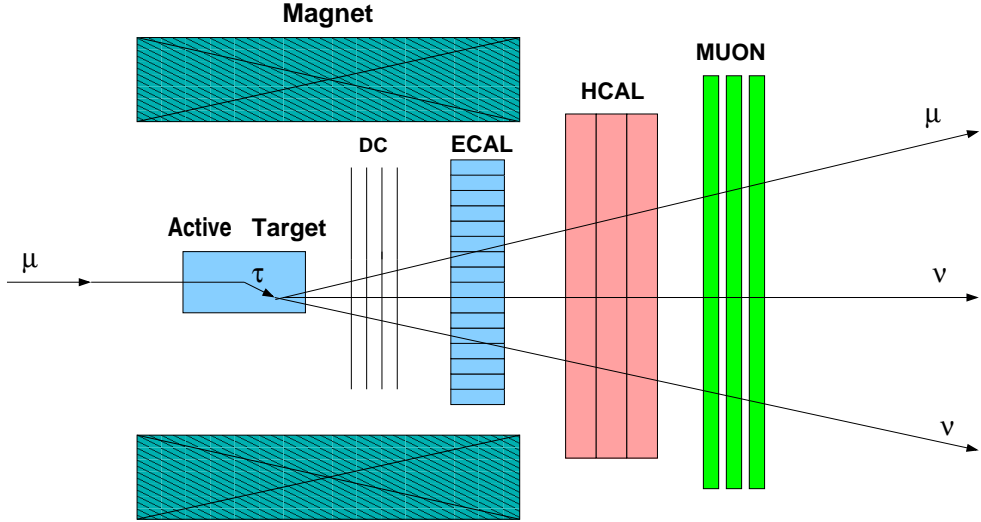,width=130mm}}
\vspace{1.cm}
 \caption{\em Schematic illustration of the "missing energy" type experiment 
on production of $\tau$ via $\mu \to \tau $ conversion in the active target
 and its detection via $\tau \to \mu \nu \nu$ decay. The experimental 
signature of the $\mu \to \tau$ conversion is a single muon 
in the final state with a catastrophic energy loss in the target.
The muon momentum is measured by the drift chamber (DC) 
spectrometer. The muon is accompanied by no significant activity
in the electromagnetic calorimeter (ECAL) and hadronic calorimeter (HCAL).}
\label{detector}
\end{figure}

As an example, we simulated the experiment on searching for 
 $\mu \rightarrow \tau$ conversion  performed with the
 NOMAD detector \cite{detector} at a high purity  sign-selected 20 GeV muon 
beam.\  
The experiment is based  
 on searching for single muon events from the primary muon interactions 
with a  target (see Figure \ref{detector}) with considerable  
energy losses which are not detected.\  The simulation were partly based
on the Monte Carlo  program used at NOMAD for the standard
neutrino interactions \cite{nomad}.

 The NOMAD detector, see Figure \ref{detector}, consists of
a number of sub-detectors most of which are located inside a 0.4 T dipole
magnet:
drift chambers (DC) with an average density of $0.1~\rm g/cm^3$ and a total 
thickness of about one radiation length ($\sim 1.0 X_0$)
 followed by a transition radiation detector and a preshower detector 
not shown in Figure \ref{detector},  and a lead-glass electromagnetic
calorimeter (ECAL).\ A hadron calorimeter (HCAL) and muon stations
are located just after the magnet coils.\ For simplicity, we have assumed  
that the primary muon interactions occur in a fully active 
dense target which is a block of lead glass,  
so that the energy losses in the target are measured.\

The sensitivity to $\mutau$ conversion and  background level were studied
for events with a simple topology of the final state.\ Namely,
 we consider quasi-elastic (QE) $\mutau$ conversion $\mu +N \to \tau + N$
at a single nucleon $N$ with the subsequent $\tau$ decay in the target.\ 
For deep inelastic events the level of background was found to be
considerably higher.\\

 Consider the  $\tau \to \mu \nu \nu$ decay of $\tau$'s produced in 
the $\mutau$ conversion.\ Since  $\tau$-leptons are very short lived 
particles - even at energy of $\simeq$ 100 GeV the average decay length of 
$\tau$ lepton is of the order of a millimetre - practically 
they are not detectable.\
The experimental signature of their presence in the beam
would be a signal of fractional  "disappearance" of primary muon beam 
energy in the detector.\footnote{Another possible way of $\tau$ identification is based on the 
algorithms developed by NOMAD for searches for $\tau$'s 
produced through $\nu_{\mu} \to \nu_{\tau}$ oscillations \cite{nomad}.}
 Indeed, two neutrinos from 
the  $\tau \to \mu \nu \nu$ decay would penetrate any type of calorimeter
without significant attenuation  and cannot be
observed effectively in the detector via its interactions.\
Hence, the only visible energy is the energy associated with the decay muons.
The experimental signature for  the  $\mu \rightarrow \tau$ conversion 
in the detector is a single muon of the same sign 
 and  with energy  less than that from the primary muon beam, 
Figure \ref{detector}.

\begin{figure}
\centering
{\hspace{-1.5cm} \epsfig{file=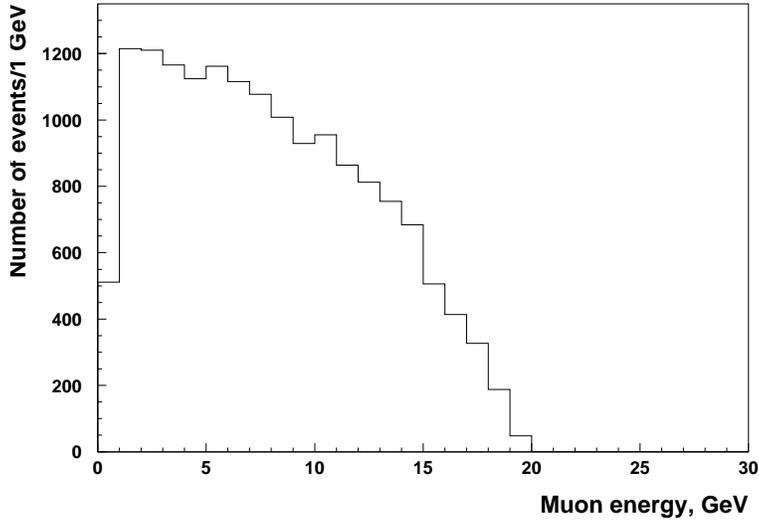,width=100mm}}
\vspace{1.cm}
 \caption{\em Simulated muon energy distribution from 
$\tau \to \mu \nu \nu$ decay of
$\tau$'s produced in  $\mu \to \tau $ conversion at energy $E_{\mu}=20$ GeV.}
\label{spectrum}
\end{figure}

The energy distribution of muons from $\taumu$ decays of $\tau$'s 
produced in the $\mutau$ conversion is shown in Figure \ref{spectrum}.\
The total efficiency of single muon detection is about 11\%
and dominated by the branching ration of tau to  muon decay mode of 17 \%
and by the muon momentum cuts.\

There are several potential sources of the background for the experiment 
considered:
\begin{itemize}
\item low energy tail in the muon beam energy distribution, e.g. due  
  presence of pions in the beam and their decays in flight;
\item catastrophic muon energy loss in the target with poorly
detected products of the reaction, e.g. deep inelastic muon scattering 
accompanied by neutrons or nuclei break-up;
\item the detector is not hermetic in the forward direction;
\item muon missidentification.
\end{itemize}  

The largest contribution to the background is expected from muon 
inelastic photonuclear interactions in the target yielding a muon and  
neutral penetrating particles in the final state (e.g. neutrons, $K^0_L$, 
..) which are not detected, e.g due to the fact that the detector is not 
fully hermetic, thus making hermiticity to be the crucial parameter. 
The effect of leakages depends on beam energy. 
 According to the simulation, the probability for secondaries
 to be mismeasured decreases with the muon energy increasing. 
Additional way to improve the sensitivity is to use active target with a low 
threshold and detectors with rather good energy resolution capability.\
At higher energies when the $\tau$-decay length 
$L_{\tau}\gtrsim$1 mm, the active target could be an instrumented target, 
e.g. silicon micro-strip detector, to measure a non zero impact parameter 
 of the secondary tracks from $\tau$ decays with respect to the main
vertex \cite{jj}. This technique being combined with the ``missing energy''
measurements will dramatically improve the sensitivity of the 
$\mu \to \tau$ conversion search. Note that the target could be the existing
STAR detector (instrumented Silicon Target) which was built and installed 
in the NOMAD detector \cite{star}.

\begin{figure}[-h]
\centering
{\hspace{-1.5cm} \epsfig{file=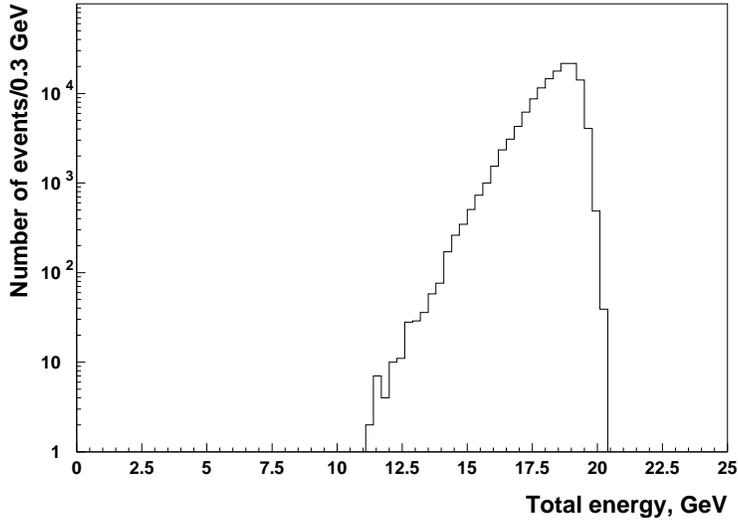,width=100mm}}
\vspace{1.cm}
 \caption{\em Distribution of total reconstructed energy in the detector 
 muon-nucleon inelastic interactions.}
\label{energy}
\end{figure}

The distribution of the reconstructed energy for 
muon-nucleon inelastic reactions  
 is shown in Figure \ref{energy}.\   
 The muon produced in the  $\mu \to \taumu$ chain is defined as a single track 
accompanied by no activity in others subdetectors by using the 
 following selection criteria:

\begin{itemize}
\item energy deposition in the target is consistent with the one 
deposited by a going through muon;
\item single DC track 
with momentum $3 < P < 10$ GeV/c matched to the single track in the 
MUON stations;
\item  no $\gamma$'s in the ECAL with energy $E_{\gamma}>0.4 ~\rm GeV$;
\item no HCAL activity: total HCAL energy $<1.4~\rm GeV$.\ This cut serves as
 an HCAL veto;
\end{itemize}


\begin{figure}[-h]
\centering
{\hspace{-1.5cm} \epsfig{file=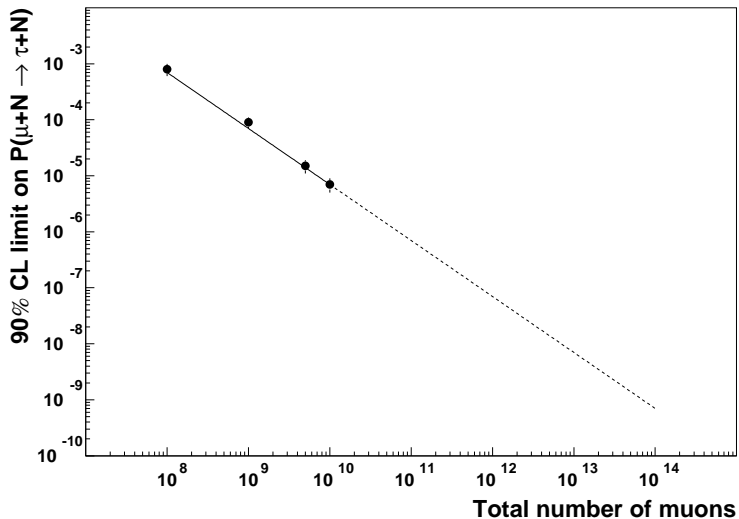,width=100mm}}
\vspace{1.cm}
 \caption{\em Simulation: 90 \% $CL$ limit on probability of the $\mutau$
conversion in the target calculated  as a function 
of total number of muons passing through the target.  
Solid line is a polynomial fit to the Monte Carlo
points, dashed line is its extrapolation to the region of higher
 number of muons.}
\label{sensitivity}
\end{figure}

 The probability $P(\mu \to \tau)$ for $\mutau$ conversion and the  number 
$N(\mutau \to \mu \nu\nu)$ of signal events in the detector 
are related by  
\begin{equation}
P(\mu \to \tau) = 
\frac{\sigma(\mu N \rightarrow \tau  N)}
{\sigma_{in}(\mu N \rightarrow  \mu ...)}
\end{equation}
and 
\begin{equation}
N(\mutau \to \mu \nu\nu)= N_{\mu}\cdot \sigma(\mu N\to \tau N)\cdot 
BR(\tau \to \mu \nu \nu)\cdot \epsilon \cdot \rho \cdot L \cdot N_A
\end{equation}
where  $N_{\mu}$ is the total number of muons, 
$\sigma_{in}(\mu N \rightarrow  \mu ...)\simeq 3 \mu b$ for the energy 
range discussed \cite{geant} 
and $\sigma(\mu N \to \tau N)$ are  the 
muon-nucleon inelastic cross section and cross section of
$\mu \to \tau$ conversion per nucleon, respectively\footnote{For simplicity
we neglect the difference between cross sections on proton and neutron,
shadow effect, etc..},
$BR(\tau \to \mu \nu \nu)$  is the branching ratio for $\tau \to \mu \nu \nu$
decay mode, $\epsilon$ is overall detection efficiency, $\rho,~L$ are 
target related factors  and $N_A$ is the Avogadro number.\

Figure \ref{sensitivity} shows estimated 
 the $90\%~CL$ upper limit for the probability 
of $\mu \to \tau$ conversion calculated based on the expected number of 
background events as a function of the total number of muons passing
through the target. 
The study of the background in the region of high sensitivity 
($P(\mu \to \tau) \lesssim 10^{-6}$)
requires  simulation of a very 
large sample of events resulting in a prohibitively large amount of computer
time.\ The maximum number of simulated 
events in our sample was about $10^6$ events. Assumption that the 
extrapolation  shown in Figure \ref{sensitivity} is valid 
down to $P(\mu \to \tau)\simeq 10^{-9}$ is probably too optimistic, however 
seems has some promise.
More detail study of the simulated background shape
is needed with significantly increased Monte Carlo statistics.\ In addition,
the question how reliable are Monte Carlo predictions in the region of
$P(\mu \rightarrow \tau) \lesssim 10^{-9}$ has also to be studied.\\

We note that direct experimental searches for a signal from
 the  $\mu \rightarrow \tau$ conversion can be performed, for example, in the
framework of the COMPASS experiment  at CERN \cite{compass}.\
For muon beam energy region $E_{\mu} >$30 GeV used in COMPASS the advantage
is that the muon-nucleon inelastic cross section increases and  
 the sensitivity to the  $\mu \rightarrow \tau$ conversion  rate 
might be expected to be of the order of magnitude high than shown in 
Figure \ref{sensitivity}.  The search for flavour changing reaction 
$\mu N \rightarrow \tau + ...$ will allow to test flavour 
changing interactions  at the level $10^{-2}G$ in terms of the corresponding 
Fermi coupling.

\section{Conclusion}
In this note, we  derive 
 phenomenological bounds on the strength of  
$\bar{\tau}\nu_{\mu}\bar{q}q^{'}$ four fermion flavour 
changing  interactions.\  Such 
interactions can arise in models with leptoquarks, R-parity violating 
supersymmetry and additional flavour changing Higgs boson interactions.\

The bounds are extracted from the combined result of 
NOMAD and CHORUS 
experiments on search for $\nu_{\mu} \to \nu_{\tau}$ oscillations, which 
constraints the rate of the $\tau$-lepton production in 
 $\nu_{\mu}N$ scattering.\ Some of these bounds supersede limits from 
rare decays. 

We also propose to 
probe  flavour changing interactions by searching 
 $\mu \to \tau$ conversion on nucleons in a
``missing energy'' type  experiment. This experiment could be performed 
at present accelerator, e.g. at high energy at CERN in the framework of the 
COMPASS experiment \cite{compass} or at the future neutrino factory.\
The instrumented target could be the STAR detector which has been installed 
and successfully tested in the NOMAD detector \cite{star}.
The estimate shows that one
might expect a sensitivity to the probability of 
 $\mu \to \tau$ conversion on nucleons to be of the order of $10^{-9}$
or even better.\ 

{\large \bf Acknowledgements}

We thank L. Camilleri and L. Di Lella for useful discussions.

\end{document}